\newcommand*{\ATLASLATEXPATH}{./}
\documentclass[preprint,texmf,texlive=2011,UKenglish,txfonts,siunitx=false]{atlasdoc}
\pdfoutput=1
\usepackage{atlasphysics}
\usepackage[biblatex=false,subfigure=true]{atlaspackage}
\usepackage{multirow}
\usepackage{cite}
\usepackage{bm}

\newcommand\tauhad{$\tau_{\text{had}}$}

\newcommand\mll{\ensuremath{m_{\ell\ell'}}}
\newcommand\Msnutau{\ensuremath{m_{\tilde{\nu}_{\tau}}}}
\newcommand\snutau{\ensuremath{\tilde{\nu}_{\tau}}}

\newcommand{\vmet}{\ensuremath{\vec{\slash\kern-.7emE}_{T}}}

\newcommand{\msnt}{\ensuremath{m_{\tilde{\nu}_{\tau}}}}

\newcommand{\metpaul}{\mbox{$\rlap{\kern0.15em/}E_T$}}

\graphicspath{{logos/}}

\AtlasTitle{Search for a Heavy Neutral Particle Decaying to $e\mu$, $e\tau$, or $\mu\tau$ in $pp$ Collisions at $\sqrt{s}=8$ TeV with the ATLAS Detector}

\PreprintIdNumber{CERN-PH-EP-2015-028}

\AtlasJournal{Physical Review Letters}

\AtlasAbstract{ %
This Letter presents a search for a heavy neutral particle decaying into an opposite-sign different-flavor
dilepton pair, $e^\pm \mu^\mp$, $e^\pm \tau^\mp$, or $\mu^\pm \tau^\mp$ using
20.3 fb$^{-1}$ of $pp$ collision data at $\sqrt{s} = 8$~TeV collected by the ATLAS detector at the LHC. The numbers of observed
candidate events are compatible with the Standard Model expectations.
Limits are set on the cross section of new phenomena in two scenarios:
the production of $\tilde{\nu}_{\tau}$ in $R$-parity-violating supersymmetric models and the production of a lepton-flavor-violating $Z'$ vector boson.
}

\begin{document}

Lepton-flavor-violation (LFV) in the charged sector, if observed at present sensitivities, 
would be a clear signature of new physics.
Such signatures occur in several new physics scenarios,
including $R$-parity-violating (RPV)~\cite{rpvsusy} supersymmetry (SUSY)~\cite{susy1,susy2,susy3,susy4,susy5,susy6,susy7,susy8,susy9} 
and models with an additional heavy neutral gauge boson $Z'$~\cite{zprm} allowing for LFV couplings. 

In RPV SUSY, the lagrangian terms allowing LFV can be expressed as $\frac{1}{2} \lambda_{ijk} L_iL_j\bar{e}_k + \lambda'_{ijk}L_iQ_j\bar{d}_k$~\cite{rpvsusy}, 
where $L$ and $Q$ are the $SU(2)$ doublet superfields of leptons and quarks; $e$ and $d$ are the $SU(2)$ singlet superfields of leptons and down-like quarks; 
$\lambda$ and $\lambda'$ are Yukawa couplings; and the indices $i$, $j$ and $k$ denote fermion generations.
A $\tau$ sneutrino ($\tilde{\nu}_\tau$) may be produced in $pp$ collisions by $d\bar{d}$
annihilation and subsequently decay to $e\mu$, $e\tau$, or $\mu\tau$.
Although only $\tilde{\nu}_{\tau}$ is considered here in order to compare with 
previous searches performed at the Tevatron, the results of our analysis apply to any 
sneutrino flavor. 

The Sequential Standard Model (SSM), where the $Z'$ 
boson is often assumed to have the same quark and lepton couplings as the Standard
Model (SM) Z boson, can be extended to include LFV couplings for the $Z'$. 
The $Z' \rightarrow e\mu$, $e\tau$, or $\mu\tau$ couplings ($Q_{12}$, $Q_{13}$, or $Q_{23}$)~\cite{zprm2} 
are typically expressed as fractions of the 
SSM $Z' \rightarrow \ell^+ \ell^-$ ($\ell=e, \mu, \tau$) coupling.

The CDF~\cite{mycdf}, D0~\cite{myd0}, and ATLAS~\cite{myatlas}
collaborations have searched for a $\tilde{\nu}_{\tau}$ in LFV final states and 
placed limits for various $\tilde{\nu}_{\tau}$ mass hypotheses. 
Both the CDF~\cite{cdfzp} and ATLAS~\cite{atlaszp} 
collaborations have placed limits on $Q_{12}$ as a function of the $Z'$ mass. 

This Letter describes 
a search for a neutral heavy particle ($\tilde{\nu}_\tau$ or $Z'$) decaying into 
$e^\pm \mu^\mp$ ($e\mu$), $e^\pm \tau_{\text{had}}^\mp$ ($e\tau$),
or $\mu^\pm \tau_{\text{had}}^\mp$ ($\mu \tau$) 
using $pp$ collision data collected at $\sqrt{s} = 8$ TeV, where \tauhad\ is a $\tau$ lepton 
that decays into hadrons. The ATLAS detector is described in detail elsewhere~\cite{atlas}.
Events are selected with a three-level trigger system that requires one or two leptons ($e$ or $\mu$) 
with high transverse momentum (\pt). 
The dataset has a total integrated luminosity of $20.3\pm 0.6$~fb$^{-1}$, where the uncertainty is derived 
following the same methodology as that detailed in Ref. \cite{lumi}.

Electrons are required to have $\pt > 25$~GeV, $|\eta| < 1.37$
or $1.52<|\eta|<2.47$~\cite{coordinate}, and
satisfy the ``\texttt{tight}'' selection in Ref. \cite{eid}, which was modified in 2012 to reduce the 
impact of additional inelastic $pp$ interactions, termed pileup.
Muon candidates must have $\pt > 25$~GeV, $|\eta|<2.4$ and be reconstructed in both the inner tracker detector and the muon spectrometer. 
The muon momenta measured by the inner detector and muon spectrometer must match
within five standard deviations of their combined uncertainty.
Good quality reconstruction and $\pt$\ resolution at high momentum are ensured by requiring a minimum number 
of associated hits on the inner detector track~\cite{muonreco} and in each of the three muon spectrometer 
stations.

Candidate events must contain at least one primary interaction vertex reconstructed with
more than three associated tracks with $\pt >$ 400 MeV. If there is more than one such vertex, 
the one with the highest sum of $\pt^{2}$ of associated tracks is chosen. 
The longitudinal impact parameter is required to be smaller than 2 mm for candidate electrons and smaller 
than 1 mm for candidate muons. It is further required that 
the transverse impact parameter is less than six times its resolution 
for candidate electrons, and that the transverse impact parameter is smaller 
than 0.2 mm for candidate muons. 
A calorimeter isolation criterion $\et^{\Delta R<0.2}/\et<0.06$ and a tracker 
isolation criterion $\pt^{\Delta R<0.4}/\pt<0.06$ are applied for both the electrons and muons, 
where $\et^{\Delta R<0.2}$ is the transverse energy 
deposited in the calorimeter within a cone of size 
$\Delta R = \sqrt{(\Delta \eta)^2+(\Delta \phi)^2} = 0.2$ around the lepton, and $\pt^{\Delta R<0.4}$ is 
the sum of the $\pt$\ of tracks with $\pt>1$ GeV within a cone size of 0.4 around the lepton. 
$\et$\ and $\pt$\ are the lepton transverse energy and momentum, respectively. 

Hadronic decays of $\tau$ leptons are characterized by one or three charged tracks associated to a narrow energy cluster in the calorimeters~\cite{atlastau}.
This search uses \tauhad\ candidates with only one charged track due to reduced 
identification and reconstruction efficiency for
high-$\pt$ $\tau$ decays with three charged tracks.  
Boosted-decision-tree multivariate discriminators, based on tracking and calorimeter
information, are used to reject jets and electrons misidentified as \tauhad.
The \tauhad\ candidates must have $|\eta| < 2.47$ and $\et > 25$ GeV. 
Candidates with $|\eta|<0.03$ are removed to exclude a region with 
increased misidentification of electrons due to reduced coverage by the inner detector and calorimeters. 

Jets are reconstructed from clusters of energy in the calorimeter using the anti-$k_t$ 
algorithm~\cite{antikt} with radius parameter $R = 0.4$. 
Jet energies are calibrated using Monte Carlo (MC) simulation and a combination of several in situ 
calibrations~\cite{myjet}. 
The calculation of the missing transverse momentum vector $\vec{p}_{\text T}^{\hspace{1.5pt}\text {miss}}$
(with magnitude \met) is based on 
the vector sum of the calibrated $\pt$\ of reconstructed jets,
electrons, and muons, as well as calorimeter energy clusters not associated with 
reconstructed objects~\cite{met}.

Candidate signal events are required to have exactly two leptons, of opposite charge and of different flavor, 
satisfying the above lepton selection criteria. 
The two leptons are required to be back-to-back in the azimuthal plane with 
$|\Delta \phi_{\ell\ell'}|>2.7$, where $\Delta \phi_{\ell\ell'}$ is the $\phi$ difference between 
the two leptons. 
Due to the presence of the undetected neutrino in the $\tau$ decay, 
the $\et$\ of the \tauhad\ candidate is required to be less than the $\et$\ ($\pt$) of the electron (muon) 
for the $e$\tauhad\ ($\mu$\tauhad) channel.

A collinear neutrino approximation is used to reconstruct the dilepton invariant mass (\mll) in the $e$\tauhad\ and $\mu$\tauhad\ channels. 
In the hadronic decay of a $\tau$ lepton from a heavy resonance, the neutrino and the resultant jet are 
nearly collinear. 
The four-vector of the neutrino is reconstructed from the $\vec{p}_{\text T}^{\hspace{1.5pt}\text {miss}}$
and $\eta$ of the \tauhad\ jet.
Four-vectors of the electron or muon, \tauhad\ candidate and neutrino are 
then used to calculate \mll. 
For $e$\tauhad\ and $\mu$\tauhad\ signal events, the above technique significantly improves 
the mass resolution and search sensitivity.

Events with \mll$<200$ GeV form a validation region to verify the background modeling, and events with 
\mll$>200$ GeV are used as the search region.

The SM processes that produce $\ell^\pm \ell'^\mp$ final states can be divided into two categories: processes that produce two prompt leptons 
such as $Z/\gamma^* \rightarrow \tau\tau$, $t\bar{t}$, single-top $Wt$ channel, diboson production, 
and processes  
where one or more photons or jets are misidentified as leptons, predominantly
$W/Z+\gamma$, $W/Z+$jets, and multijet events.
The decay of a $\tau$ to an electron or a muon is considered as prompt production. 
For the $e$\tauhad~($\mu$\tauhad) channel, additional background can originate from the $Z/\gamma^* \rightarrow ee~(\mu\mu)$ 
process if one lepton is misidentified as a \tauhad\ candidate. 
The contributions of these processes are even larger with respect to the
$Z/\gamma^* \rightarrow \tau\tau$ background, since the final states e or $\mu$ are usually harder
than those from leptonic $\tau$ decay. 

Contributions from processes in the prompt two-lepton category, as well as photon-related and $Z/\gamma^* \rightarrow ee~(\mu\mu)$ backgrounds,
are estimated using MC simulation~\cite{atlassim}. The detector response model is based on the 
{\sc geant}4 program~\cite{g4}. Lepton reconstruction and identification efficiencies, energy scales, and resolutions in
the MC simulation are corrected to the corresponding values measured in the data. 
Pileup is included to match distributions observed in the data. 
Top quark production is generated with {\sc mc@nlo} v4.06~\cite{mcatnlo} 
for $t\bar{t}$ and single-top, the Drell--Yan process ($Z/\gamma^* \rightarrow \ell\ell$) is generated with {\sc alpgen} v2.14~\cite{alpgen}, and diboson processes are generated 
with {\sc herwig} v6.520.2~\cite{herwig}. 
Samples of $W\gamma$ and $Z\gamma$ events are generated with {\sc sherpa} v1.04~\cite{sherpa}.
These generated samples are normalized to the most accurate available cross-section calculations. 
For the 
dominant backgrounds, the Drell--Yan processes are corrected to 
next-to-next-to-leading order (NNLO)~\cite{nll7}, and $t\bar{t}$ is corrected to NNLO, 
including soft-gluon resummation to next-to-next-to-leading-logarithm order~\cite{ttbarxsect}. 

Since it is difficult to model misidentification of jets as leptons, particularly at high \pt,
the $W+$jets and multijet backgrounds are determined from control regions in the data.
The $W+$jets background is determined in 
a control region selected with the same criteria as used for the signal selection except requiring $\met>30$~GeV
(to enhance the $W$ contribution) and requiring that 
the electron or muon \pt\ be less than 150~GeV (to eliminate potential signal).
Simulation studies indicate that there is negligible multijet background in this control region.
For the $e$\tauhad\ and $\mu$\tauhad\ channels, the number of events in the control region is
corrected for the other SM background sources using MC samples.
For the $e\mu$ channel, the number of $W+$jets events in the control region is too small to yield a 
statistically meaningful measurement.
Instead, the control region is enlarged by removing the 
isolation criterion on one lepton, 
and the $W+$jets contribution is estimated using 
the lepton $\et^{\Delta R<0.2}/\et$ distribution to fit the data with the MC predictions 
for other SM processes (dominant at low values of the isolation variable) 
and $W+$jets (dominant at large values).
For the $W+$jets background in all three channels, the extrapolation 
factor from the control region to the signal region and the shape of the \mll\ distribution are 
taken from the $W+$jets MC sample.

The contribution from multijet production is estimated from a control region with the same 
selection as the signal region except that the leptons are required to have the same electric charge, under 
the assumption that the probability of misidentifying a jet as a lepton is independent of the charge.
The number of events in this same-sign control region is corrected for contributions from backgrounds
with prompt leptons and from $W+$jets backgrounds using the procedure described above.
The assumption of charge independence of the jet misidentification rates is tested in 
a multijet-enriched region with two nonisolated leptons.  
After subtraction of other SM backgrounds, the ratio of opposite-sign to same-sign events is 
found to be $1.07 \pm 0.06~({\text{stat.}}) \pm 0.02~({\text{syst.}})$.

The background estimates are verified in validation regions defined with the same selection criteria 
as for the signal regions but with $1.0<|\Delta \phi_{\ell\ell'}|<2.7$.  In these validation regions, 
simulation studies show the backgrounds have compositions similar to the signal regions, and the
predictions agree with the data within 20\% over the entire \mll\ range. 
An uncertainty of 20\% is hence placed on the total background estimate 
that is used in the final results. 

MC signal events are generated with {\sc herwig} v6.520.2 for $\tilde{\nu}_{\tau}$ and {\sc pythia}
v8.165~\cite{pythia} for $Z'$.
Samples are produced with $\tilde{\nu}_{\tau}$ and $Z'$ masses ranging from 0.5 to 3~TeV. 
Signal cross sections are calculated to next-to-leading order for $\tilde{\nu}_{\tau}$, 
and NNLO for $Z'$.
The theoretical uncertainties are taken from an envelope of cross-section predictions using
different parton distribution function (PDF) sets and factorization and renormalization scales~\cite{nll6,zpxsect}.


The selection efficiency, including $\tau$ decay branching ratio if $\tau$ is involved,  
for $m_{\tilde{\nu}_{\tau}}$ = 2  TeV are 42\%, 14\%, and 10\% 
in the $e\mu$, $e\tau$ and $\mu\tau$ channels, respectively. 
The corresponding numbers for a $Z'$ boson with $m_{Z'}$ = 2 TeV are 37\%, 11\%, and 9\%.  
The systematic uncertainties on the signal efficiency 
vary from 3\% to 6\% depending on the resonance mass and 
decay mode.  The primary contributions are due to the number of MC events, and 
the uncertainties related to the muon and $\tau$ \pT\ scales. 

\begin{figure*}[htb]
\begin{center}
\includegraphics[width=0.32\textwidth]{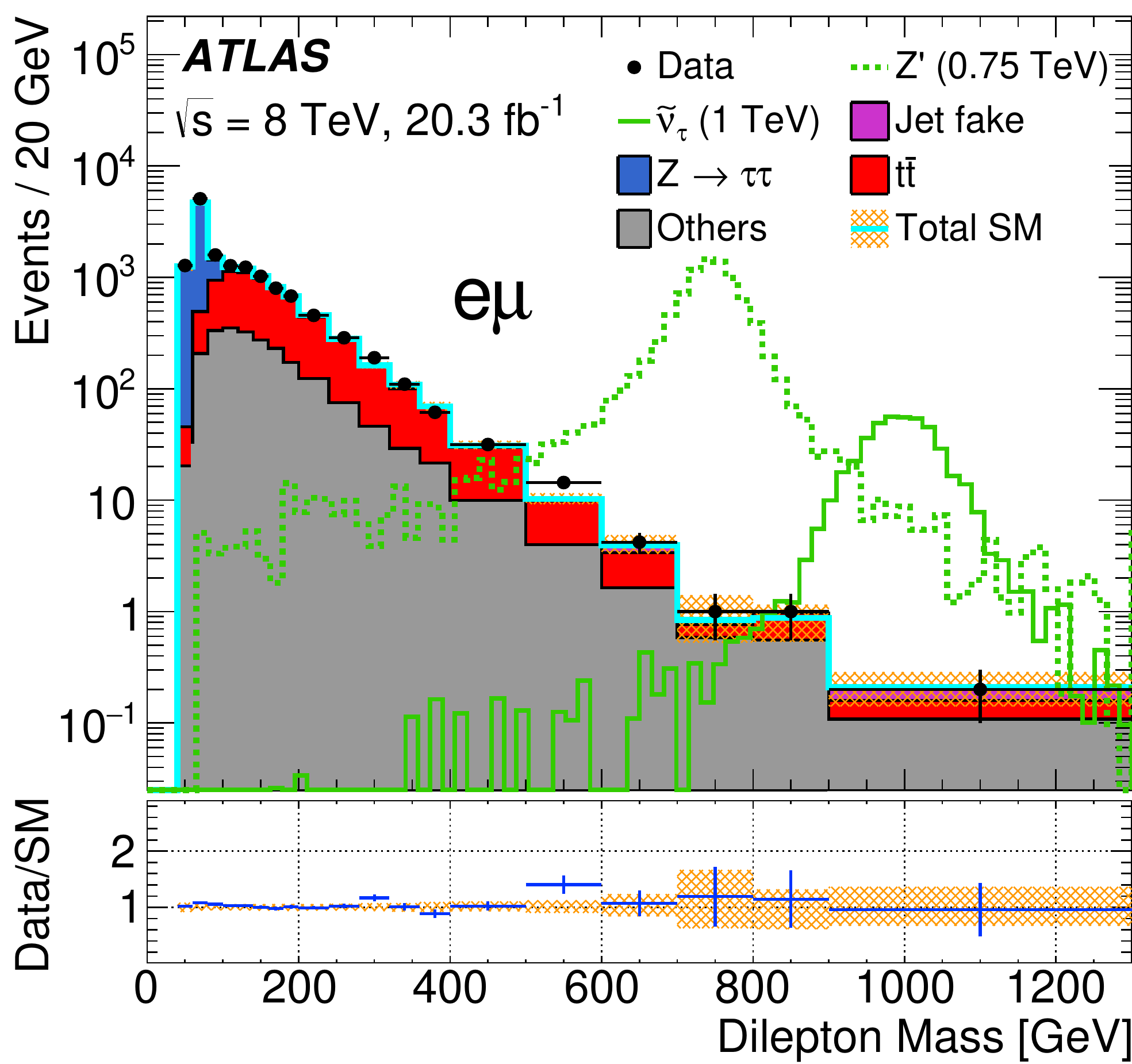}
\includegraphics[width=0.32\textwidth]{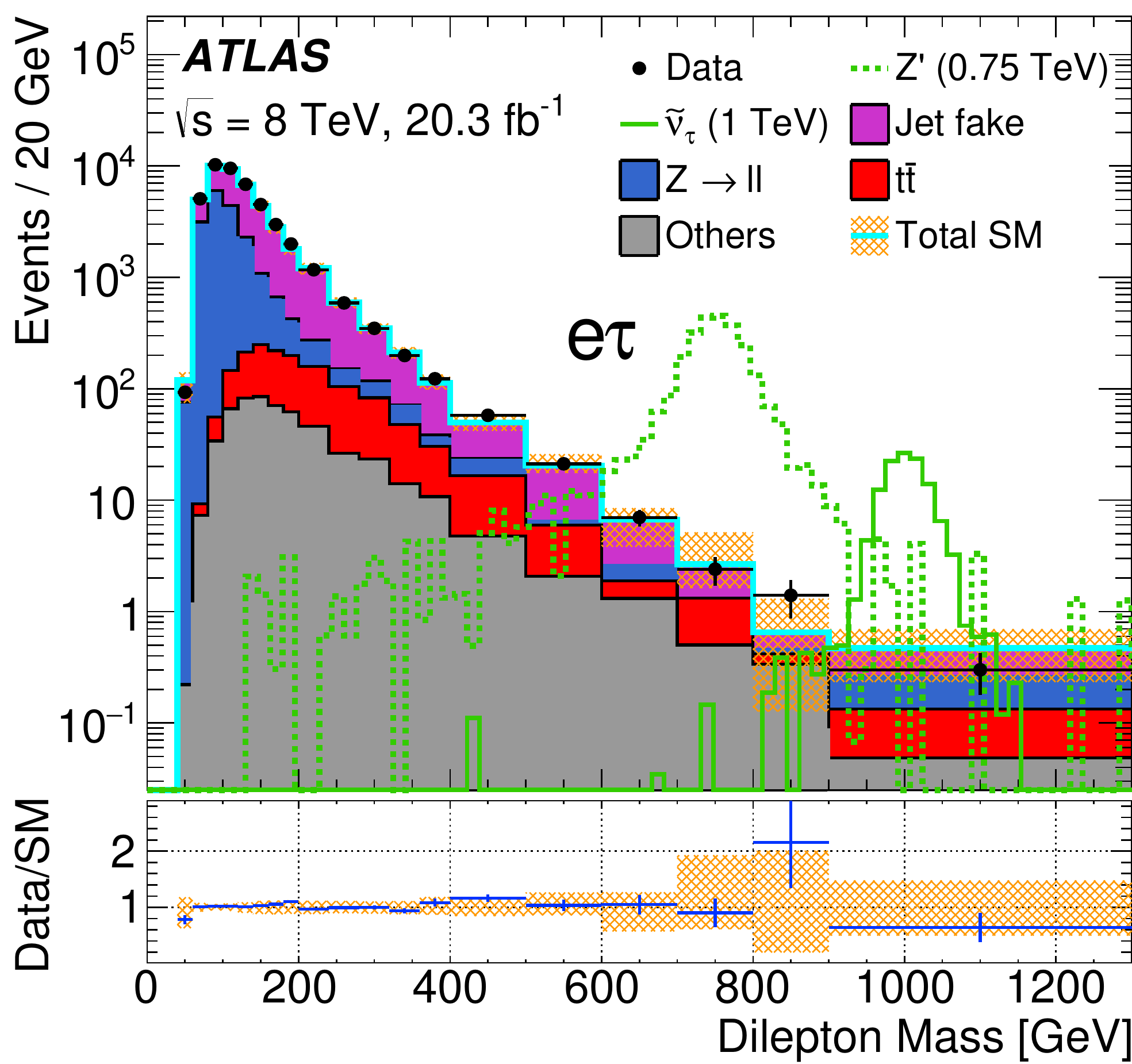}
\includegraphics[width=0.32\textwidth]{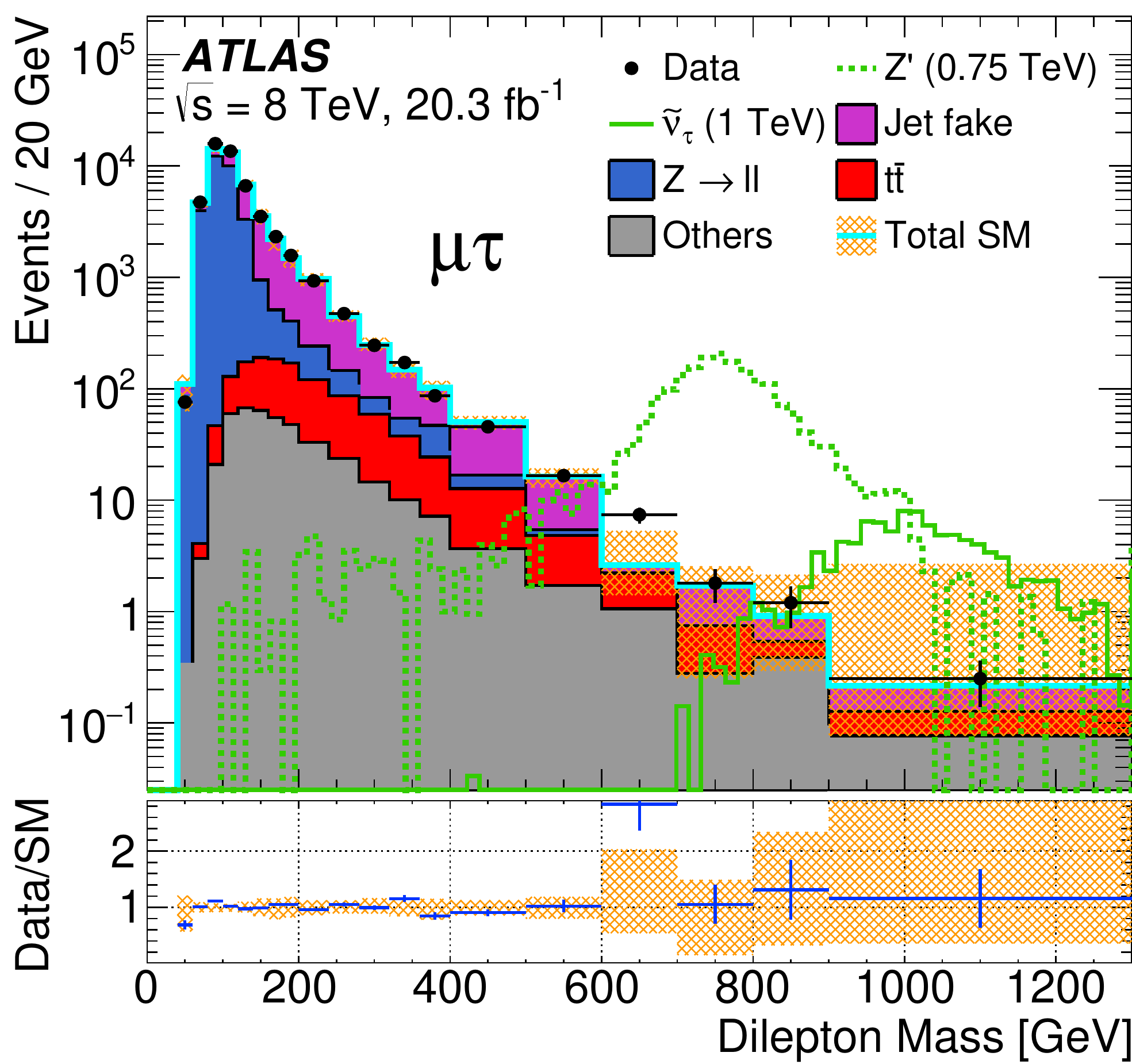}
\caption{Observed and predicted $e\mu$, $e$\tauhad, $\mu$\tauhad\ invariant mass 
distributions. The contributions of the different processes are also shown: ``Others'' 
includes diboson and single-top while ``Jet fake'' refers to $W$+jets and multijet. 
All overflows are included in the rightmost bin. 
Signal simulations are shown for \Msnutau$=1$ TeV and $m_{Z'}=0.75$ TeV. The 
couplings $\lambda'_{311}=0.11$ and $\lambda_{i3k}=0.07$ ($Q_{\ell\ell'}= 1$) are used for the 
RPV ($Z'$) model. The uncertainty bands include both the statistical and systematic uncertainties.}
\label{fig:mll}
\end{center}
\end{figure*}

The observed and expected event yields in both the validation and search \mll\ regions 
for all three final states are in good agreement, as summarized 
in Table~\ref{tab:summary}. 
The \mll\ distributions (Fig.~\ref{fig:mll}) 
show no significant excess above the SM expectation in any of the three modes. The dominant contributions 
to the uncertainty bands in Fig.~\ref{fig:mll} are due to the number of MC events, 
the MC cross-section uncertainties,  the \met\ scale and resolution, and the uncertainty in
the shape of the \mll\ distribution for $W+$jets. 

\newcommand{\pho}{\phantom{0}}
\begin{table*}
\begin{center}
\caption{\label{tab:data_mc_cutflow} Estimated SM backgrounds and observed event yields for
the validation ($m_{\ell \ell^{\prime}} < 200$~GeV) and search ($m_{\ell \ell^{\prime}} > 200$~GeV) regions.
Both the statistical and systematic uncertainties are included.
Due to correlations, the total uncertainties
are not exactly the sum in quadrature of the components.}
\label{tab:summary}
\begin{tabular}{@{} l|r@{$\pm$}rr@{$\pm$}rr@{$\pm$}r|r@{$\pm$}rr@{$\pm$}rr@{$\pm$}r}
\hline
    & \multicolumn{6}{c}{$m_{\ell \ell^{\prime}} < 200$~GeV} & \multicolumn{6}{|c}{$m_{\ell \ell^{\prime}} > 200$~GeV}  \\
\hline
  Process & \multicolumn{2}{c}{$N_{e \mu}$} & \multicolumn{2}{c}{$N_{e \tau_\mathrm{had}}$} & \multicolumn{2}{c}{$N_{\mu \tau_\mathrm{had}}$} & \multicolumn{2}{|c}{$N_{e \mu}$} & \multicolumn{2}{c}{$N_{e \tau_\mathrm{had}}$} & \multicolumn{2}{c}{$N_{\mu \tau_\mathrm{had}}$} \\ \hline
 $Z/\gamma^* \rightarrow \tau \tau$    & 6000&400 & 11000&900  & 11200&700  & 28&12    & 72&21   & 99&33    \\
 $Z/\gamma^* \rightarrow e e$          &  \multicolumn{2}{c}{---}    & 6100&1100  & \multicolumn{2}{c|}{---}  & \multicolumn{2}{c}{---}   & 430&70  & \multicolumn{2}{c}{---}    \\
 $Z/\gamma^* \rightarrow \mu \mu$      &  \multicolumn{2}{c}{---}    &  \multicolumn{2}{c}{---}      & 19500&1300 &  \multicolumn{2}{c}{---}    & \multicolumn{2}{c}{---}   & 410&80   \\
 $t\bar{t}$                            & 4220&290 & 690&60     & 580&50     & 1640&120 & 700&60  & 550&40   \\
 Diboson                               &  1440&80 &  321&29    &  258&17    &  474&30  &  197&17 &  141&11     \\
 Single top quark                      &  470&40  &  87&11     &  60&7      &  202&17    &  90&10    &  73&8      \\
  $W$+jets                              &   54&18  &  17000&4000   &  14000&4000   &  8&4   &  3600&700  &  2800&600  \\
 Multijet                              &   227&32 &  4800&1000   &  700&800    &  58&12     &  340&210  &  100&190     \\ \hline
  Total                           & 12400&600 & 40400&2900 & 46000&4000 & 2400&130 & 5400&500 & 4200&400  \\
\hline
  Data                            & \multicolumn{2}{c}{12954}  & \multicolumn{2}{c}{41304}   & \multicolumn{2}{c}{48304}   &\multicolumn{2}{|c}{2474} & \multicolumn{2}{c}{5336}& \multicolumn{2}{c}{4184}      \\
\hline
\end{tabular}
\end{center}
\end{table*}

Upper limits are placed on the production cross section times branching ratio [$\sigma(pp \rightarrow \tilde{\nu}_{\tau}/Z') \times \textrm{BR}(\tilde{\nu}_{\tau}/Z' \rightarrow \ell \ell^{\prime})$].
For each $\tilde{\nu}_\tau$ or $Z'$ mass, $m$, the search region is defined to be 
$m \pm\ 3\sigma_{\ell\ell'}$, 
where $\sigma_{\ell\ell'}$ is the standard deviation of the simulated signal \mll\ distribution.
The relative width of the signal \mll\ distribution ranges from 3\% to 17\% for different mass points, 
channels and models. 
To reduce the statistical error, if the upper side of the search region is greater than 1 TeV, all events above 1 TeV are used. 
To further reduce the effect of fluctuations in the high-mass region due to low MC event counts, the number of 
background events in each mass window is estimated using a double exponential fit to the total background \mll\ distribution.
The fit uncertainty is taken into account in the limit-setting procedure, including a 
contribution from varying the fit function range. 

A frequentist technique~\cite{freq} is used to set the expected and observed upper limits as a function of \Msnutau\ and $m_{Z'}$. 
The likelihood of observing the number of events in data as a function of the expected number of signal 
and background events is constructed from a Poisson distribution for each \mll\ bin. 
Systematic uncertainties are taken into account with Gaussian-distributed nuisance parameters.
A 95\% confidence level (CL) limit is then determined. 
The expected exclusion limits are determined, using simulated pseudoexperiments containing only background
processes, as the median of the 95\% CL limit distributions for 
each set of pseudoexperiments at each 
value of $\msnt$ or $m_{Z'}$, including systematic uncertainties.
The ensemble of limits is also used to assess the $1\sigma$ and $2\sigma$ uncertainty envelopes
of the expected limits. 

\begin{figure*}[htb]
\begin{center}
   \includegraphics[width=0.85\textwidth,height=220pt]{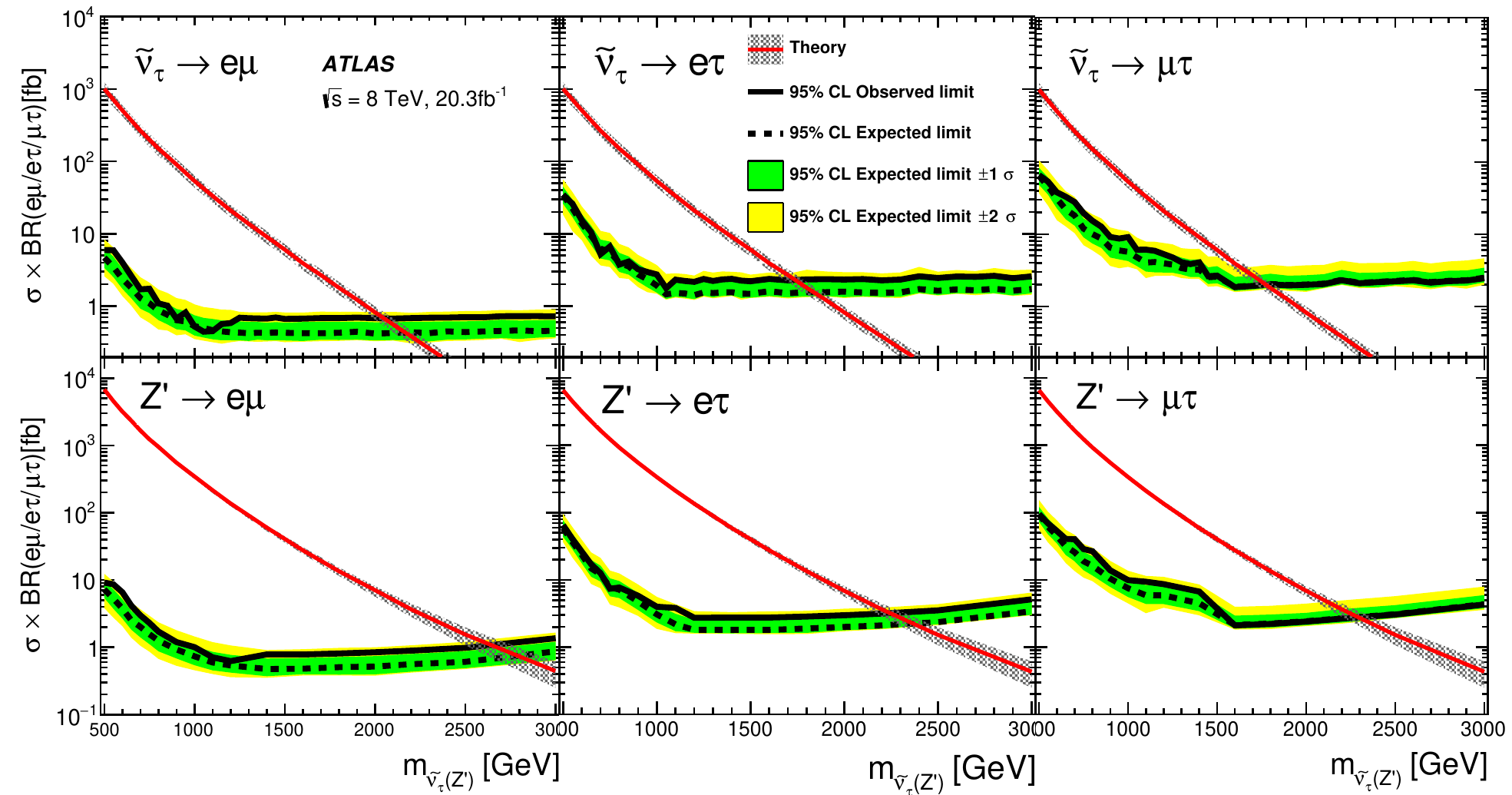}
\caption{ The 95\% CL limits on cross section times branching ratio as a function of $\tilde{\nu}_\tau$ 
mass (top plots) and $Z'$ mass (bottom plots) for $e\mu$ (left), $e\tau$ (middle), and $\mu\tau$ (right).  
Theory curves are for the arbitrary choice of couplings $\lambda'_{311}=0.11$ and $\lambda_{i3k}=0.07$ for 
$\tilde{\nu}_{\tau}$ and 
$Q_{\ell\ell'} = 1$ for $Z'$. The gray band around the theory curve represents the theoretical
uncertainty from the PDFs and factorization and renormalization scales.
}
\label{fig:coupling_limits}
\end{center}
\end{figure*}

Figure~\ref{fig:coupling_limits} shows the observed and expected cross section times branching ratio 
limits as a function of $\msnt$($m_{Z'}$), 
together with the $\pm 1\sigma$ and $\pm 2\sigma$ uncertainty bands. 
For a $\tilde{\nu}_{\tau}$ mass of 1~TeV, the observed limits on the production cross section 
times branching ratio are 0.5 fb, 2.7 fb, and 9.1 fb for the $e\mu$, $e\tau$ and $\mu\tau$ channels, 
respectively. The corresponding limits for a $Z'$ boson mass of 1~TeV are 1.0 fb, 4.0 fb and 9.9 fb for 
the $e\mu$, $e\tau$ and $\mu\tau$ channels, respectively. 

Theoretical predictions of cross section times branching ratio~\cite{stnxsect,nll7} are also shown, 
assuming $\lambda'_{311}=0.11$ and $\lambda_{i3k}=0.07$ for 
the $\tilde{\nu}_{\tau}$  and $Q_{ij} = 1$ for the $Z'$, consistent with benchmark 
couplings used in previous searches. 

For these benchmark couplings, the lower limits on the \snutau\  mass 
are 2.0 TeV, 1.7 TeV, and 1.7 TeV for the $e \mu$, $e \tau$ 
and $\mu \tau$ channels, respectively. The corresponding lower limits on the $Z'$ mass are 
2.5 TeV, 2.2 TeV and 2.2 TeV for the $e \mu$, $e \tau$ and $\mu \tau$ channels, respectively. 
The observed lower mass limits are a factor of three to four higher than the best limits from 
the Tevatron~\cite{mycdf,myd0} and also more stringent than 
the previous limits 
from ATLAS~\cite{myatlas} for the same couplings. 

In summary, a search has been performed for a heavy particle decaying to $e \mu$, 
$e \tau_\mathrm{had}$, or $\mu \tau_\mathrm{had}$ final states using 20.3~fb$^{-1}$ of 
$pp$ collision data at $\sqrt{s}=8$ TeV recorded by the \mbox{ATLAS} detector at the LHC. 
The data are found to be consistent with SM predictions. 
Limits are placed on the cross section times branching ratio for an RPV SUSY $\tilde{\nu}_\tau$ 
and a LFV $Z'$ boson. 
These results considerably extend previous constraints from the Tevatron and LHC experiments. 

We thank CERN for the very successful operation of the LHC, as well as the
support staff from our institutions without whom ATLAS could not be
operated efficiently.

We acknowledge the support of ANPCyT, Argentina; YerPhI, Armenia; ARC,
Australia; BMWFW and FWF, Austria; ANAS, Azerbaijan; SSTC, Belarus; CNPq and FAPESP,
Brazil; NSERC, NRC and CFI, Canada; CERN; CONICYT, Chile; CAS, MOST and NSFC,
China; COLCIENCIAS, Colombia; MSMT CR, MPO CR and VSC CR, Czech Republic;
DNRF, DNSRC and Lundbeck Foundation, Denmark; EPLANET, ERC and NSRF, European Union;
IN2P3-CNRS, CEA-DSM/IRFU, France; GNSF, Georgia; BMBF, DFG, HGF, MPG and AvH
Foundation, Germany; GSRT and NSRF, Greece; RGC, Hong Kong SAR, China; ISF, MINERVA, GIF, I-CORE and Benoziyo Center, Israel; INFN, Italy; MEXT and JSPS, Japan; CNRST, Morocco; FOM and NWO, Netherlands; BRF and RCN, Norway; MNiSW and NCN, Poland; GRICES and FCT, Portugal; MNE/IFA, Romania; MES of Russia and ROSATOM, Russian Federation; JINR; MSTD,
Serbia; MSSR, Slovakia; ARRS and MIZ\v{S}, Slovenia; DST/NRF, South Africa;
MINECO, Spain; SRC and Wallenberg Foundation, Sweden; SER, SNSF and Cantons of
Bern and Geneva, Switzerland; NSC, Taiwan; TAEK, Turkey; STFC, the Royal
Society and Leverhulme Trust, United Kingdom; DOE and NSF, United States of
America.

The crucial computing support from all WLCG partners is acknowledged
gratefully, in particular from CERN and the ATLAS Tier-1 facilities at
TRIUMF (Canada), NDGF (Denmark, Norway, Sweden), CC-IN2P3 (France),
KIT/GridKA (Germany), INFN-CNAF (Italy), NL-T1 (Netherlands), PIC (Spain),
ASGC (Taiwan), RAL (UK) and BNL (USA) and in the Tier-2 facilities
worldwide.

\clearpage

\begin{flushleft}
{\Large The ATLAS Collaboration}

\bigskip

G.~Aad$^{\rm 85}$,
B.~Abbott$^{\rm 113}$,
J.~Abdallah$^{\rm 152}$,
O.~Abdinov$^{\rm 11}$,
R.~Aben$^{\rm 107}$,
M.~Abolins$^{\rm 90}$,
O.S.~AbouZeid$^{\rm 159}$,
H.~Abramowicz$^{\rm 154}$,
H.~Abreu$^{\rm 153}$,
R.~Abreu$^{\rm 30}$
% ...
\bigskip
\\
% ...
$^{11}$ Institute of Physics, Azerbaijan Academy of Sciences, Baku, Azerbaijan\\
% ...
$^{30}$ CERN, Geneva, Switzerland\\
% ...
$^{85}$ CPPM, Aix-Marseille Universit{\'e} and CNRS/IN2P3, Marseille, France\\
% ...
$^{90}$ Department of Physics and Astronomy, Michigan State University, East Lansing MI, United States of America\\
% ...
$^{107}$ Nikhef National Institute for Subatomic Physics and University of Amsterdam, Amsterdam, Netherlands\\
% ...
$^{113}$ Homer L. Dodge Department of Physics and Astronomy, University of Oklahoma, Norman OK, United States of America\\
% ...
$^{152}$ Institute of Physics, Academia Sinica, Taipei, Taiwan\\
$^{153}$ Department of Physics, Technion: Israel Institute of Technology, Haifa, Israel\\
$^{154}$ Raymond and Beverly Sackler School of Physics and Astronomy, Tel Aviv University, Tel Aviv, Israel\\
% ...
$^{159}$ Department of Physics, University of Toronto, Toronto ON, Canada\\
% ...
$^{*}$ Deceased
\end{flushleft}


\begin{thebibliography}{00}
\bibitem{rpvsusy}
R. Barbier {\it et al.}, Phys. Rep. {\bf 420}, 1 (2005), [\href{http://arxiv.org/abs/hep-ph/0406039}{arXiv:hep$-$ph/0406039}].

\bibitem{susy1} 
H. Miyazawa, Prog. Theor. Phys. {\bf 36} (6), 1266 (1966).

\bibitem{susy2} 
P. Ramond, Phys. Rev. D {\bf 3}, 2415 (1971).

\bibitem{susy3} 
Y. A. Golfand and E. P. Likhtman, JETP Lett. {\bf 13}, 323 (1971), 
[Pisma Zh. Eksp. Teor. Fiz. 13:452,1971].

\bibitem{susy4} 
A. Neveu and J. H. Schwarz, Nucl. Phys. B {\bf 31}, 86 (1971).

\bibitem{susy5}  
A. Neveu and J. H. Schwarz, Phys. Rev. D {\bf 4}, 1109 (1971).

\bibitem{susy6}  
J. Gervais and B. Sakita, Nucl. Phys. B {\bf 34}, 632 (1971).

\bibitem{susy7}  
D. V. Volkov and V. P. Akulov, Phys. Lett. B {\bf 46}, 109 (1973).

\bibitem{susy8}  
J. Wess and B. Zumino, Phys. Lett. B {\bf 49}, 52 (1974).

\bibitem{susy9}  
J. Wess and B. Zumino, Nucl. Phys. B {\bf 70}, 39 (1974).

\bibitem{zprm}
P. Langacker, Rev. Mod. Phys. {\bf 81}, 1199 (2009), [\href{http://arxiv.org/abs/0801.1345}{arXiv:0801.1345}].

\bibitem{zprm2}
B. Murakami, Phys. Rev. D {\bf 65}, 055003 (2002), [\href{http://arxiv.org/abs/hep-ph/0110095}{arXiv:hep$-$ph/0110095}].

\bibitem{mycdf}
T. Aaltonen {\it et al.} (CDF Collaboration),
Phys. Rev. Lett. {\bf 105}, 191801 (2010), [\href{http://arxiv.org/abs/1004.3042}{arXiv: 1004.3042}]; A. Abulencia {\it et al.} (CDF Collaboration), Phys. Rev. Lett. {\bf 96}, 211802 (2006), [\href{http://arxiv.org/abs/hep-ex/0603006}{arXiv:hep$-$ex/0603006}].

\bibitem{myd0}
V. M. Abazov {\it et al.} (D0 Collaboration), Phys. Rev. Lett. {\bf 105}, 191802 (2010), [\href{http://arxiv.org/abs/1007.4835}{arXiv:1007.4835}]; V. M. Abazov {\it et al.} (D0 Collaboration), Phys. Rev. Lett. {\bf 100}, 241803 (2008), [\href{http://arxiv.org/abs/0711.3207}{arXiv:0711.3207}].

\bibitem{myatlas}
ATLAS Collaboration, Phys. Lett. B {\bf 723}, 15 (2013), [\href{http://arxiv.org/abs/1212.1272}{arXiv:1212.1272}].

\bibitem{cdfzp}
T. Aaltonen {\it et al.} (CDF Collaboration), Phys. Rev. Lett. {\bf 105}, 191801 (2010), [\href{http://arxiv.org/abs/1004.3042}{arXiv:1004.3042}].

\bibitem{atlaszp}
ATLAS Collaboration, Eur. Phys. J. C {\bf 71}, 1809 (2011),  [\href{http://arxiv.org/abs/1109.3089}{arXiv:1109.3089}].

\bibitem{atlas}
ATLAS Collaboration, JINST {\bf 3}, S08003 (2008),  [\href{http://arxiv.org/abs/0901.0512}{arXiv:0901.0512}].

\bibitem{lumi}
ATLAS Collaboration, Eur. Phys. J. C {\bf 73}, 2518 (2013), [\href{http://arxiv.org/abs/1302.4393}{arXiv:1302.4393}].

\bibitem{coordinate}
ATLAS uses a right-handed coordinate system with its origin at the nominal interaction 
point in the center of the detector and the $z$-axis along the beam direction. The $x$-axis points toward the center of 
the LHC ring, and the $y$-axis points upward. Cylindrical coordinates $(r,\phi)$ are used in the transverse plane, $\phi$ being 
the azimuthal angle around the beam direction. Pseudorapidity is defined in terms of the polar angle $\theta$ as 
$\eta=-\ln\tan(\theta/2)$. Transverse projections are defined relative to the beam axis. 

\bibitem{eid}
ATLAS Collaboration, Eur. Phys. J. C {\bf 74}, 2941 (2014), [\href{http://arxiv.org/abs/1404.2240}{arXiv:1404.2240}].

\bibitem{muonreco}
ATLAS Collaboration, Eur. Phys. J. C {\bf 74}, 3130 (2014), [\href{http://arxiv.org/abs/1407.3935}{arXiv:1407.3935}].

\bibitem{atlastau}
ATLAS Collaboration, ATLAS-CONF-2013-064, https://cds.cern.ch/record/1562839 .


\bibitem{antikt}
M. Cacciari, G. P. Salam, and G. Soyez, J. High Energy Phys. 04 (2008) 063, [\href{http://arxiv.org/abs/0802.1189}{arXiv:0802.1189}].

\bibitem{myjet}
ATLAS Collaboration, Eur. Phys. J. C {\bf 73}, 2304 (2013), [\href{http://arxiv.org/abs/1112.6426}{arXiv:1112.6426}].

\bibitem{met} 
ATLAS Collaboration, Eur.\ Phys.\ J.\ C {\bf 72}, 1844 (2012), [\href{http://arxiv.org/abs/1108.5602}{arXiv:1108.5602}].


\bibitem{atlassim}
ATLAS Collaboration, Eur. Phys. J. C {\bf 70}, 823 (2010), [\href{http://arxiv.org/abs/1005.4568}{arXiv:1005.4568}].

\bibitem{g4}
S. Agostinelli {\it et al.}, Nucl. Instrum. Methods Phys. Res., Sect. A {\bf 506}, 250 (2003).

\bibitem{mcatnlo}
S. Frixione and B.R.Webber, J. High Energy Phys. 06 (2002) 029, [\href{http://arxiv.org/abs/hep-ph/0204244}{arXiv:hep$-$ph/0204244}]; S. Frixione, E. Laenen and P. Motylinski, J. High Energy Phys. 03 (2006) 092, [\href{http://arxiv.org/abs/hep-ph/0512250}{arXiv:hep$-$ph/0512250}]; S. Frixione {\it et al.}, J. High Energy Phys. 07 (2008) 029, [\href{http://arxiv.org/abs/0805.3067}{arXiv:0805.3067}].

\bibitem{alpgen}
M.L.Mangano, M.Moretti, F.Piccinini, R.Pittau, A.Polosa, J. High Energy Phys. 07 (2003) 001, [\href{http://arxiv.org/abs/hep-ph/0206293}{arXiv:hep$-$ph/0206293}].

\bibitem{herwig}
G. Corcella {\it et al.}, J. High Energy Phys. 01 (2001) 010, [\href{http://arxiv.org/abs/hep-ph/0011363}{arXiv:hep$-$ph/0011363}]. 


\bibitem{sherpa}
 T. Gleisberg {\it et al.},  J. High Energy Phys. 02 (2009) 007, [\href{http://arxiv.org/abs/0811.4622}{arXiv:0811.4622}]; T.~Gleisberg and S.~Hoeche, J. High Energy Phys. 12 (2008) 039, [\href{http://arxiv.org/abs/0808.3674}{arXiv:0808.3674}]; S.~Schumann and F.~Krauss,  J. High Energy Phys. 03 (2008) 038, [\href{http://arxiv.org/abs/0709.1027}{arXiv:0709.1027}]; S.~Hoeche, F.~Krauss, S.~Schumann and F.~Siegert, J. High Energy Phys. 05 (2009) 053, [\href{http://arxiv.org/abs/0903.1219}{arXiv:0903.1219}].

\bibitem{nll7}
Y. Li and F. Petriello, Phys. Rev. D {\bf 86}, 094034 (2012), [\href{http://arxiv.org/abs/1208.5967}{arXiv:1208.5967}].

\bibitem{ttbarxsect}
M. Czakon, P. Fiedler, and A. Mitov,  Comput. Phys. Commun.{\bf 185}, 2930 (2014), [\href{http://arxiv.org/abs/1112.5675}{arXiv:1112.5675}]; M. Czakon and A. Mitov, Phys. Rev. Lett. {\bf 110}, 252004 (2013), [\href{http://arxiv.org/abs/1303.6254}{arXiv:1303.6254}. 

\bibitem{pythia}
T. Sj\"{o}strand, S. Mrenna, and P. Skands, J. High Energy Phys. 05 (2006) 026, [\href{http://arxiv.org/abs/hep-ph/0603175}{arXiv:hep-ph/0603175}]; T. Sj\"{o}strand, S. Mrenna, and P. Skands, Comput. Phys. Comm. {\bf 178}, 852 (2008), [\href{http://arxiv.org/abs/0710.3820}{arXiv:0710.3820}].

\bibitem{nll6}
M. Kramer {\it et al.}, \href{http://arxiv.org/abs/1206.2892}{[arXiv:1206.2892]}.

\bibitem{zpxsect}
ATLAS Collaboration, Phys. Rev. D {\bf 90}, 052005 (2014), [\href{http://arxiv.org/abs/1405.4123}{arXiv:1405.4123}].

\bibitem{freq}
G. Cowan {\it et al.}, Eur. Phys. J. C {\bf 71}, 1554 (2011), [\href{http://arxiv.org/abs/1007.1727}{arXiv:1007.1727}].

\bibitem{stnxsect}
S. Wang {\it et al.}, Chin. Phys. Lett. {\bf 25}, 58 (2008), [\href{http://arxiv.org/abs/0706.3079}{arXiv:0706.3079}].

\end{thebibliography}
\end{document}